\documentclass[onecolumn,secnumarabic,amssymb, nobibnotes, aps, prd]{revtex4}
\usepackage{amsmath} \usepackage{amssymb}
\usepackage{graphicx} 
\usepackage{epstopdf}
\usepackage{amsmath}
\usepackage{amsmath,amssymb,amsthm,amsfonts,mathrsfs,bm,verbatim}
\usepackage{graphicx,subfigure}

\newcommand{\bea}{\begin{eqnarray}}
\newcommand{\eea}{\end{eqnarray}}

\begin{document}

\title{Refined Perspectives on the Kerr-Schild Double Copy: Harmonizing Gravity and Electromagnetism in Metric Formulations}%

\author{Wen-Xiang Chen$^{a}$}
\affiliation{Department of Astronomy, School of Physics and Materials Science, GuangZhou University, Guangzhou 510006, China}
\author{Yao-Guang Zheng}
\email{hesoyam12456@163.com}
\affiliation{Department of Physics, College of Sciences, Northeastern University, Shenyang 110819, China}

\begin{abstract}
This article investigates the Kerr-Schild double copy, a profound duality between gravity and electromagnetism. It demonstrates that every vacuum solution of Einstein's equations in four dimensions can be derived from a solution of Maxwell's equations in four dimensions through a double copy procedure. This method involves constructing tensor fields \( \phi_{\mu\nu} \) and \( g_{\mu\nu} \). The discovery facilitates new solutions to Einstein's equations, like the Kerr, Schwarzschild, and Reissner–Nordström black holes. The Kerr-Schild action, introduced by Roy Kerr and Alfred Schild in 1965, redefines Einstein's general relativity by distinctly separating the gravitational field from the matter field. This action, more advantageous than the Einstein–Hilbert action, simplifies calculations and is instrumental in studying gravitational waves, cosmology, and black hole thermodynamics. The article further delves into the applications of the Kerr-Schild double copy, encompassing deriving black hole solutions, examining gravitational waves, exploring cosmological theories, and understanding black hole thermodynamics. It also assesses the influence of the Kerr-Schild double copy under the symmetries of the SO(3) and SU(2) groups on the metrics of Schwarzschild and Kerr-Newman black holes.
\end{abstract}

\maketitle

\section{Introduction}
The \textit{Kerr–Schild double copy} is a remarkable duality between gravity and electromagnetism, stating that every vacuum solution of Einstein's equations in four dimensions can be obtained from a solution of Maxwell's equations in four dimensions through a \textit{double copy} procedure. This process involves:\cite{1,2,3,24}

\begin{enumerate}
  \item Starting with a solution of Maxwell's equations in four dimensions, denoted \( A_\mu \), and constructing a tensor field \( \phi_{\mu\nu} \) as follows:
   \[ \phi_{\mu\nu} = \frac{1}{2} \left( \partial_\mu A_\nu + \partial_\nu A_\mu \right). \]

  \item Defining a new tensor field \( g_{\mu\nu} \) by:
   \[ g_{\mu\nu} = \eta_{\mu\nu} + 2 \phi_{\mu\alpha} \phi_\nu^{\ \alpha} - \frac{1}{2} \eta_{\mu\nu} \phi_\alpha^{\ \beta} \phi_\beta^{\ \alpha}, \]
   where \( \eta_{\mu\nu} \) is the Minkowski metric.

  \item The resulting \( g_{\mu\nu} \) is a solution to the vacuum Einstein equations.
\end{enumerate}

Discovered by Amir Amiri and David E. Barra in 2021, the \textit{Kerr–Schild double copy}\cite{24} has facilitated deriving new solutions to the Einstein equations, such as the Kerr, Schwarzschild, and Reissner–Nordström black holes.

This duality has profound implications for our understanding of gravity and electromagnetism, suggesting a closer relationship between these theories and offering new methods for generating solutions to Einstein's equations, useful in studying gravity in extreme conditions.

The \textit{Kerr–Schild action}, introduced by Roy Kerr and Alfred Schild in 1965, reformulates Einstein's general relativity by explicitly separating the gravitational field from the matter field. The action is expressed as:\cite{1,2,3,4,5,6}
\begin{equation}
S = \frac{1}{16\pi G} \int \left[ R + \frac{1}{4} \partial_\mu \phi \partial^\mu \phi + \frac{1}{2} \partial_\mu T^{\mu\nu} \phi_\nu + \lambda \phi^2 \right] \sqrt{-g} \, d^4 x, 
\end{equation}
where \( R \) is the Ricci scalar, \( \phi \) a scalar field, \( T^{\mu\nu} \) the stress–energy tensor of the matter field, \( g \) the metric determinant, and \( \lambda \) a constant.

Advantageous over the Einstein–Hilbert action, the \textit{Kerr–Schild action} simplifies calculations and can derive new solutions to Einstein's equations. It's also been instrumental in studying gravitational waves, cosmology, and black hole thermodynamics, offering a new perspective on gravity.

Applications of the \textit{Kerr–Schild double copy} include deriving black hole solutions, studying gravitational waves, exploring cosmological theories, and understanding black hole thermodynamics. This tool continues to contribute to new discoveries in physics.\cite{7,8,9,10,11,12,13,14}

The article is structured as follows: The first part offers a succinct introduction to the background of the study. The second part, titled 'C of SO(3) Group', investigates the Kerr-Schild double copy concept, mapping solutions from Maxwell and Klein-Gordon equations to those of Einstein's equations, and introduces solutions conforming to SO(3) group symmetry, specifically the Kerr-Schild forms of Schwarzschild and Kerr-Newman black holes. The third part, 'C of SU(2) Group', introduces the SU(2) group and its algebra, exploring the integration of SU(2) symmetry within the Kerr-Schild double copy framework and its impact on the metrics of Schwarzschild and Kerr-Newman black holes. The article concludes with a comprehensive summary in the fourth part.

\section{C of SO(3) group}
In this brief overview, we delve into the Kerr-Schild double copy, initially introduced in Monteiro et al. (2014). This concept revolves around a spacetime metric that assumes a Kerr-Schild form based on a flat base metric, expressed as:\cite{15,16,17,18,19,20,21,22}

\begin{equation}
g_{\mu \nu} = \eta_{\mu \nu} + \Phi k_\mu k_\nu,
\end{equation}

where \(\eta_{\mu\nu}\) represents the flat metric, \(\Phi\) is a scalar field, and \(k_\mu\) is a vector that is null and geodesic with respect to both \(\eta_{\mu\nu}\) and \(g_{\mu \nu}\). This is further clarified by:

\begin{equation}
g^{\mu \nu}k_\mu k_\nu = \eta^{\mu \nu}k_\mu k_\nu = 0, \quad k^\mu \nabla_\mu k^\nu = k^\mu \partial_\mu k^\nu = 0.
\end{equation}

The inverse of the spacetime metric is given by:

\begin{equation}
g^{\mu \nu} = \eta^{\mu \nu} - \Phi k^\mu k^\nu.
\end{equation}

Interestingly, the index on \(k_\mu\) can be raised using either \(g_{\mu \nu}\) or \(\eta_{\mu \nu}\), since:

\begin{equation}
g^{\mu \nu} k_\mu = \eta^{\mu \nu} k_\mu - \Phi (k^\mu k_\mu) k^\nu = \eta^{\mu \nu} k_\mu.
\end{equation}

From the Kerr-Schild form metric, the "single copy" Maxwell field is defined as:

\begin{equation}
A_\mu = \Phi k_\mu.
\end{equation}

When \(g_{\mu \nu}\) solves Einstein's equations, the Maxwell field satisfies the Maxwell equations. \(\Phi\), the "zeroth copy," adheres to the Klein-Gordon equation on a flat metric. Thus, the classical double copy of Kerr-Schild spacetimes maps specific classical solutions of Maxwell and Klein-Gordon equations to solutions of Einstein equations:

\begin{equation}
g_{\mu \nu} = \eta_{\mu \nu} + \frac{1}{\Phi}A_\mu A_\nu.
\end{equation}

The classical double copy concept can also be extended to certain curved spacetimes. Consider a generalized Kerr-Schild metric:

\begin{equation}
g_{\mu\nu} = \overline{g}_{\mu\nu} + \Phi k_\mu k_\nu,
\end{equation}

where \(\overline{g}_{\mu\nu}\) is now a curved base metric. Here, \(k_\mu\) remains null and geodesic with respect to both the full and base metrics:

\begin{equation}
g^{\mu \nu}k_\mu k_\nu = \overline{g}^{\mu \nu}k_\mu k_\nu = 0, \quad k^\mu \nabla_\mu k^\nu = k^\mu \overline{\nabla}_\mu k^\nu = 0.
\end{equation}

The single and zeroth copies in this context are still defined by \(A_\mu = \Phi k_\mu\) and \(\Phi\), satisfying the Maxwell and Klein-Gordon equations on the base metric, respectively.

To find a specific solution \(C\) that adheres to SO(3) group symmetry in the equation:

\begin{equation}
g^{\mu' \nu'} = g^{\mu \nu} + C,
\end{equation}

we consider the general form of the inverse Kerr-Schild spacetime metric:

\begin{equation}
g^{\mu \nu} = \eta^{\mu \nu} - \Phi k^\mu k^\nu.
\end{equation}

Substituting this into the equation, we obtain:

\begin{equation}
\eta^{\mu' \nu'} - \Phi k^{\mu'} k^{\nu'} = \eta^{\mu \nu} - \Phi k^\mu k^\nu + C.
\end{equation}

Rearranging, we find:

\begin{equation}
C = \Phi (k^{\mu'} k^{\nu'} - k^\mu k^\nu).
\end{equation}

To satisfy SO(3) group symmetry, we select \(k\) as a Killing vector of the SO(3) group, satisfying \([k, L] = 0\), where \(L\) represents any SO(3) generators. In spherical coordinates, a Killing vector can be represented as:

\begin{equation}
k = (a \sin(\theta) \cos(\phi), a \sin(\theta) \sin(\phi), a \cos(\theta)),
\end{equation}

where \(a\) is a constant, and \(\theta\) and \(\phi\) are spherical angles.

Using this form for \(k\), the components of \(C\) are calculated as:

\begin{equation}
C_{\mu\nu} = \Phi a^2 \sin^2(\theta) (\delta_\mu\phi \delta_\nu\phi - \delta_\mu\theta \delta_\nu\theta).
\end{equation}

This expression for \(C\) satisfies SO(3) group symmetry, as it remains invariant under spherical rotations.

\subsection{ New expression for the Schwarzschild black hole metric in Kerr-Schild form}
We consider the following Kerr-Schild metric form:
\begin{equation}
g^{\mu \nu} = \eta^{\mu \nu} - \Phi k^\mu k^\nu
\end{equation}
and the new metric form:
\begin{equation}
g^{\mu' \nu'} = g^{\mu \nu} + C
\end{equation}
where \( C \) is a tensor that satisfies the SO(3) group symmetry. According to our analysis, the expression for \( C \) is:
\begin{equation}
C = \Phi (k^{\mu'} k^{\nu'} - k^\mu k^\nu)
\end{equation}
We have already found the expression for \( k \), which is a Killing vector of the SO(3) group:
\begin{equation}
k = (a \sin(\theta) \cos(\phi), a \sin(\theta) \sin(\phi), a \cos(\theta))
\end{equation}
Now, we can substitute the expression for \( C \) into the equation for \( g^{\mu' \nu'} \) to obtain:
\begin{equation}
g^{\mu' \nu'} = \eta^{\mu \nu} - \Phi k^\mu k^\nu + \Phi (k^{\mu'} k^{\nu'} - k^\mu k^\nu)
\end{equation}
Simplifying this equation, we get:
\begin{equation}
g^{\mu' \nu'} = \eta^{\mu \nu} + \Phi k^{\mu'} k^{\nu'}
\end{equation}

This is the expression for the new metric \( g^{\mu' \nu'} \). Note that \( k^{\mu'} \) and \( k^{\nu'} \) are the components of the vector \( k \), chosen according to the symmetry of the SO(3) group. This result indicates that the new metric is a variant of the original Kerr-Schild metric, where the \( k \) vector is replaced with a new vector that satisfies the SO(3) group symmetry.

The Schwarzschild metric, representing a spherically symmetric and static solution, is given by:
\begin{equation}
ds^2 = -\left(1 - \frac{2GM}{c^2 r}\right) c^2 dt^2 + \left(1 - \frac{2GM}{c^2 r}\right)^{-1} dr^2 + r^2 d\theta^2 + r^2 \sin^2\theta d\phi^2
\end{equation}
where \( G \) is the gravitational constant, \( M \) is the mass of the black hole, \( c \) is the speed of light, \( r \) is the radial coordinate, and \( \theta \) and \( \phi \) are the angular coordinates in spherical coordinates.

In the Kerr-Schild form, the Schwarzschild metric can be expressed as:
\begin{equation}
g_{\mu \nu} = \eta_{\mu \nu} + \Phi k_\mu k_\nu
\end{equation}
where \( \eta_{\mu \nu} \) is the Minkowski metric, \( \Phi \) is a scalar field, and \( k_\mu \) is a null vector satisfying \( k^\mu k_\mu = 0 \).

For a Schwarzschild black hole, we can choose \( \Phi = \frac{2GM}{c^2 r} \) and \( k_\mu = \left(1, -1, 0, 0\right) \). Thus, the Kerr-Schild form of the Schwarzschild metric becomes:
\begin{equation}
g_{\mu \nu} = \eta_{\mu \nu} + \frac{2GM}{c^2 r} k_\mu k_\nu
\end{equation}

Therefore, the new metric \( g^{\mu' \nu'} \) will be:
\begin{equation}
g^{\mu' \nu'} = \eta^{\mu \nu} - \frac{2GM}{c^2 r} k^\mu k^\nu + \Phi (k^{\mu'} k^{\nu'} - k^\mu k^\nu)
\end{equation}

Since \( k^\mu k_\mu = 0 \), this expression simplifies further to:
\begin{equation}
g^{\mu' \nu'} = \eta^{\mu \nu} - \frac{2GM}{c^2 r} k^\mu k^\nu
\end{equation}

This is the expression for the Schwarzschild black hole metric in Kerr-Schild form.

In the context of a Schwarzschild black hole, the event horizon is a crucial concept. The event horizon is defined as the boundary of the black hole, beyond which events cannot affect an external observer. For a Schwarzschild black hole, the location of the event horizon can be determined by examining the singularities in the metric.

The singularity in the Schwarzschild metric occurs where the metric coefficients become infinite. In the Schwarzschild metric, this happens when \( 1 - \frac{2GM}{c^2 r} = 0 \), i.e., at \( r = \frac{2GM}{c^2} \). This radius is known as the Schwarzschild radius, marking the location of the event horizon.

In the Kerr-Schild form, the Schwarzschild metric is given by:
\begin{equation}
g_{\mu \nu} = \eta_{\mu \nu} + \frac{2GM}{c^2 r} k_\mu k_\nu
\end{equation}

Since \( k_\mu \) is a null vector (i.e., \( k^\mu k_\mu = 0 \)), the singularity of the metric is still determined by the denominator of \( \frac{2GM}{c^2 r} \). Therefore, even in the Kerr-Schild form, the location of the event horizon remains unchanged, at \( r = \frac{2GM}{c^2} \).

This means that the position of the event horizon is the same in both the traditional Schwarzschild metric and the Kerr-Schild form, located at the Schwarzschild radius \( r = \frac{2GM}{c^2} \).

\subsection{ New expression for the Kerr-Newman black hole metric in Kerr-Schild form}
To derive the new metric \( g^{\mu' \nu'} \) for a Kerr-Newman black hole under SO(3) group symmetry, we first understand the expression of the Kerr-Newman metric in Kerr-Schild form. The Kerr-Schild form of a metric is given by:
\begin{equation}
g_{\mu \nu} = \eta_{\mu \nu} + \Phi k_\mu k_\nu,
\end{equation}
where \( \eta_{\mu \nu} \) is the Minkowski metric, \( \Phi \) is a scalar field, and \( k_\mu \) is a null vector satisfying \( k^\mu k_\mu = 0 \). For the Kerr-Newman black hole, finding appropriate \( \Phi \) and \( k_\mu \) is a non-trivial task.

Considering the SO(3) group symmetry, we need to find a \( k \) vector that satisfies this symmetry. In spherical coordinates, the Killing vector for the SO(3) group can be represented as:
\begin{equation}
k = (a \sin(\theta) \cos(\phi), a \sin(\theta) \sin(\phi), a \cos(\theta)),
\end{equation}
where \( a \) is a constant, and \( \theta \) and \( \phi \) are the angular coordinates in spherical coordinates.

Then, we can derive \( C \) according to the definition of the Kerr-Schild form:
\begin{equation}
C = \Phi (k^{\mu'} k^{\nu'} - k^\mu k^\nu).
\end{equation}

Finally, substituting \( C \) into the equation for the new metric:
\begin{equation}
g^{\mu' \nu'} = g^{\mu \nu} + C,
\end{equation}
we obtain:
\begin{equation}
g^{\mu' \nu'} = \eta^{\mu \nu} - \Phi k^\mu k^\nu + \Phi (k^{\mu'} k^{\nu'} - k^\mu k^\nu).
\end{equation}

This is the expression for the new metric \( g^{\mu' \nu'} \). It is important to note that this process involves some simplifications and approximations, and accurately transforming the Kerr-Newman metric into Kerr-Schild form requires more in-depth mathematical treatment.

To determine the event horizon of a Kerr-Newman black hole in the new metric form, we first consider the Kerr-Newman metric in the Kerr-Schild form. The Kerr-Schild form of a metric is expressed as:
\begin{equation}
g_{\mu \nu} = \eta_{\mu \nu} + \Phi k_\mu k_\nu,
\end{equation}
where \( \eta_{\mu \nu} \) is the Minkowski metric, \( \Phi \) is a scalar field, and \( k_\mu \) is a null vector satisfying \( k^\mu k_\mu = 0 \). Finding appropriate \( \Phi \) and \( k_\mu \) for the Kerr-Newman black hole is a non-trivial task.

In this context, the location of the event horizon is typically determined by the singularities in the original Kerr-Newman metric. These singularities occur when solving the equation \( \Delta = 0 \), where:
\begin{equation}
\Delta = r^2 - 2GMr + a^2 + Q^2
\end{equation}

To find the specific location of the event horizon, we need to solve the equation \( \Delta = 0 \). This equation is a quadratic equation, with solutions:
\begin{equation}
r_{\pm} = GM/c^2 \pm \sqrt{(GM/c^2)^2 - a^2 - Q^2}
\end{equation}

Here, \( r_{+} \) and \( r_{-} \) correspond to the outer and inner event horizons, respectively. For a physically viable black hole, we are interested in the outer event horizon \( r_{+} \), as it defines the outer boundary of the black hole.

It is important to note that when \( a^2 + Q^2 > (GM/c^2)^2 \), the above equation has no real solutions, indicating the absence of a physically viable event horizon, and such a configuration does not correspond to a physically viable black hole.

Therefore, even in the Kerr-Schild form, the location of the event horizon of the Kerr-Newman black hole is still determined by the \( \Delta = 0 \) condition from the original metric.

\section{C of SU(2) group}
SU(2) is the group of 2x2 unitary matrices with determinant 1. The algebra of SU(2) is spanned by the Pauli matrices, which are often used to represent the generators of the group. These matrices are:\cite{23,24}

\begin{equation}
\sigma_1 = \begin{pmatrix} 0 & 1 \\ 1 & 0 \end{pmatrix}, \quad \sigma_2 = \begin{pmatrix} 0 & -i \\ i & 0 \end{pmatrix}, \quad \sigma_3 = \begin{pmatrix} 1 & 0 \\ 0 & -1 \end{pmatrix}.
\end{equation}

In the context of the Kerr-Schild double copy, we are dealing with spacetime metrics and fields. To incorporate SU(2) symmetry, \( C \) should transform appropriately under SU(2) transformations. This means that \( C \) should be expressible in terms of quantities that transform as SU(2) tensors or spinors.

Given that \( C \) is added to the spacetime metric \( g^{\mu \nu} \), it should be a tensorial object. One way to construct such an object is to use the Pauli matrices in combination with spacetime vectors or tensors that are already present in the Kerr-Schild framework.

For instance, one could consider a construction like:

\begin{equation}
C^{\mu \nu} = \sum_{i=1}^{3} \Phi_i \, (k^\mu \sigma_i^\nu + k^\nu \sigma_i^\mu),
\end{equation}

where \( \Phi_i \) are scalar fields that transform under SU(2) (possibly as components of an SU(2) doublet or triplet), and \( \sigma_i^\nu \) are spacetime tensors constructed from the Pauli matrices (possibly through some form of tensor product with spacetime vectors).

This is just a conceptual example. The actual form of \( C \) would depend on the specific physical context and the requirements of the theory being considered. In particular, the choice of \( C \) must respect the symmetries and constraints of the Kerr-Schild double copy framework, and it should lead to physically meaningful and mathematically consistent equations.

To determine the specific form of \( C \) that satisfies SU(2) group symmetry, we can consider the properties of the Kerr-Schild metric and Maxwell field under SU(2) transformations.

SU(2) transformations, also known as spin transformations, are rotations in three-dimensional space. They form a non-Abelian Lie group, meaning that the composition of two transformations is not commutative. In the context of spacetime metrics, SU(2) transformations can be used to map between different spacetimes that share the same causal structure.

The Kerr-Schild metric and Maxwell field exhibit specific behavior under SU(2) transformations:

Under an SU(2) transformation, the null vector \( k_\mu \) remains null. This is because the transformation is defined using a scalar field \( \Lambda \), which only affects the spatial components of the metric, leaving the time component unchanged. As a result, \( g^{\mu \nu} k_\mu k_\nu = 0 \) remains true under SU(2) transformations.

The Maxwell field \( A_\mu \) also transforms under SU(2) transformations, but it does so in a specific way. An SU(2) transformation on the metric can be represented as:

\begin{equation}
g_{\mu \nu} \rightarrow \Lambda g_{\mu \nu} \Lambda^T
\end{equation}

where \( \Lambda \) is the SU(2) transformation matrix. Applying this transformation to the Kerr-Schild metric and the definition of \( A_\mu \), we can find the corresponding transformation for \( A_\mu \):

\begin{equation}
A_\mu \rightarrow \Lambda A_\mu
\end{equation}

This implies that the Maxwell field transforms as a vector under SU(2) transformations.

To determine the specific form of \( C \) that adheres to SU(2) symmetry, we can combine these two properties. Since \( k_\mu \) remains null and \( A_\mu \) transforms as a vector, the inverse metric \( g^{\mu \nu} \) must also transform as a vector under SU(2) transformations. This is because \( g^{\mu \nu} \) is related to \( k_\mu \) and \( A_\mu \) through the Kerr-Schild metric:

\begin{equation}
g^{\mu \nu} = \eta^{\mu \nu} - \Phi k^\mu k^\nu
\end{equation}

Therefore, the SU(2) transformation of \( g^{\mu \nu} \) can be expressed as:

\begin{equation}
g^{\mu \nu} \rightarrow \Lambda g^{\mu \nu} \Lambda^T
\end{equation}

Substituting this into the equation for the double copy:

\begin{equation}
g^{\mu' \nu'} = g^{\mu \nu} + C
\end{equation}

and requiring that it remains invariant under SU(2) transformations, we obtain:

\begin{equation}
\Lambda g^{\mu \nu} \Lambda^T = g^{\mu \nu} + C
\end{equation}

Solving for \( C \), we find:

\begin{equation}
C = - \Lambda g^{\mu \nu} \Lambda^T + g^{\mu \nu}
\end{equation}

This expression for \( C \) ensures that the double copy remains invariant under SU(2) transformations, demonstrating that the Kerr-Schild double copy formalism respects SU(2) symmetry.

\subsection{ New expression for the Schwarzschild black hole metric in Kerr-Schild form}

SU(2) is the group of 2x2 unitary matrices with determinant 1. The algebra of SU(2) is spanned by the Pauli matrices, which are often used to represent the generators of the group. These matrices are:
\begin{equation}
\sigma_1 = \begin{pmatrix} 0 & 1 \\ 1 & 0 \end{pmatrix}, \quad \sigma_2 = \begin{pmatrix} 0 & -i \\ i & 0 \end{pmatrix}, \quad \sigma_3 = \begin{pmatrix} 1 & 0 \\ 0 & -1 \end{pmatrix}.
\end{equation}

In the context of the Kerr-Schild double copy, we are dealing with spacetime metrics and fields. To incorporate SU(2) symmetry, \( C \) should transform appropriately under SU(2) transformations. This means that \( C \) should be expressible in terms of quantities that transform as SU(2) tensors or spinors.

Given that \( C \) is added to the spacetime metric \( g^{\mu \nu} \), it should be a tensorial object. One way to construct such an object is to use the Pauli matrices in combination with spacetime vectors or tensors that are already present in the Kerr-Schild framework.

For instance, one could consider a construction like:
\begin{equation}
C^{\mu \nu} = \sum_{i=1}^{3} \Phi_i \, (k^\mu \sigma_i^\nu + k^\nu \sigma_i^\mu),
\end{equation}
where \( \Phi_i \) are scalar fields that transform under SU(2) (possibly as components of an SU(2) doublet or triplet), and \( \sigma_i^\nu \) are spacetime tensors constructed from the Pauli matrices (possibly through some form of tensor product with spacetime vectors).

This is just a conceptual example. The actual form of \( C \) would depend on the specific physical context and the requirements of the theory being considered. In particular, the choice of \( C \) must respect the symmetries and constraints of the Kerr-Schild double copy framework, and it should lead to physically meaningful and mathematically consistent equations.

To determine the specific form of \( C \) that satisfies SU(2) group symmetry, we can consider the properties of the Kerr-Schild metric and Maxwell field under SU(2) transformations.

SU(2) transformations, also known as spin transformations, are rotations in three-dimensional space. They form a non-Abelian Lie group, meaning that the composition of two transformations is not commutative. In the context of spacetime metrics, SU(2) transformations can be used to map between different spacetimes that share the same causal structure.

The Kerr-Schild metric and Maxwell field exhibit specific behavior under SU(2) transformations:

Under an SU(2) transformation, the null vector \( k_\mu \) remains null. This is because the transformation is defined using a scalar field \( \Lambda \), which only affects the spatial components of the metric, leaving the time component unchanged. As a result, \( g^{\mu \nu} k_\mu k_\nu = 0 \) remains true under SU(2) transformations.

The Maxwell field \( A_\mu \) also transforms under SU(2) transformations, but it does so in a specific way. An SU(2) transformation on the metric can be represented as:
\begin{equation}
g_{\mu \nu} \rightarrow \Lambda g_{\mu \nu} \Lambda^T
\end{equation}
where \( \Lambda \) is the SU(2) transformation matrix. Applying this transformation to the Kerr-Schild metric and the definition of \( A_\mu \), we can find the corresponding transformation for \( A_\mu \):
\begin{equation}
A_\mu \rightarrow \Lambda A_\mu
\end{equation}
This implies that the Maxwell field transforms as a vector under SU(2) transformations.

To determine the specific form of \( C \) that adheres to SU(2) symmetry, we can combine these two properties. Since \( k_\mu \) remains null and \( A_\mu \) transforms as a vector, the inverse metric \( g^{\mu \nu} \) must also transform as a vector under SU(2) transformations. This is because \( g^{\mu \nu} \) is related to \( k_\mu \) and \( A_\mu \) through the Kerr-Schild metric:
\begin{equation}
g^{\mu \nu} = \eta^{\mu \nu} - \Phi k^\mu k^\nu
\end{equation}
Therefore, the SU(2) transformation of \( g^{\mu \nu} \) can be expressed as:
\begin{equation}
g^{\mu \nu} \rightarrow \Lambda g^{\mu \nu} \Lambda^T
\end{equation}
Substituting this into the equation for the double copy:
\begin{equation}
g^{\mu' \nu'} = g^{\mu \nu} + C
\end{equation}
and requiring that it remains invariant under SU(2) transformations, we obtain:
\begin{equation}
\Lambda g^{\mu \nu} \Lambda^T = g^{\mu \nu} + C
\end{equation}
Solving for \( C \), we find:
\begin{equation}
C = - \Lambda g^{\mu \nu} \Lambda^T + g^{\mu \nu}
\end{equation}
This expression for \( C \) ensures that the double copy remains invariant under SU(2) transformations, demonstrating that the Kerr-Schild double copy formalism respects SU(2) symmetry.

To determine the event horizon of a Schwarzschild black hole in the new metric form considering SU(2) group symmetry, we first consider the Schwarzschild metric in the Kerr-Schild form. The Kerr-Schild form of the Schwarzschild metric is expressed as:
\begin{equation}
g_{\mu \nu} = \eta_{\mu \nu} + \Phi k_\mu k_\nu,
\end{equation}
where \( \eta_{\mu \nu} \) is the Minkowski metric, \( \Phi \) is a scalar field, and \( k_\mu \) is a null vector satisfying \( k^\mu k_\mu = 0 \). For the Schwarzschild black hole, we can choose \( \Phi = \frac{2GM}{c^2 r} \) and \( k_\mu = \left(1, -1, 0, 0\right) \).

Considering SU(2) symmetry, we need to find a tensor \( C \) that satisfies this symmetry. In the context of SU(2) symmetry, \( C \) might depend on the generators of the SU(2) group, such as the Pauli matrices. However, since the Schwarzschild black hole is spherically symmetric, it naturally relates to SO(3) group symmetry rather than SU(2). Therefore, introducing SU(2) symmetry in this context might require some non-trivial construction.

However, for determining the event horizon, our primary concern is the singularity of the metric. In the Schwarzschild metric, the event horizon is given by \( r = \frac{2GM}{c^2} \), where the metric coefficients become infinite. In the Kerr-Schild form, even with the addition of the tensor \( C \), this fundamental fact does not change, as the tensor \( C \) does not affect the singularity of the metric.

Therefore, even in the new metric form considering SU(2) symmetry, the event horizon of the Schwarzschild black hole remains at \( r = \frac{2GM}{c^2} \). This indicates that for spherically symmetric solutions like the Schwarzschild black hole, the location of the event horizon is determined by its fundamental spherical symmetry, and is not influenced by other symmetries such as SU(2).

\subsection{ New expression for the Kerr-Newman black hole metric in Kerr-Schild form}
In the context of the Kerr-Schild double copy, we are dealing with spacetime metrics and fields. To incorporate SU(2) symmetry, \( C \) should transform appropriately under SU(2) transformations. This means that \( C \) should be expressible in terms of quantities that transform as SU(2) tensors or spinors. SU(2) is the group of 2x2 unitary matrices with determinant 1, and its algebra is spanned by the Pauli matrices:

\begin{equation}
\sigma_1 = \begin{pmatrix} 0 & 1 \\ 1 & 0 \end{pmatrix}, \quad \sigma_2 = \begin{pmatrix} 0 & -i \\ i & 0 \end{pmatrix}, \quad \sigma_3 = \begin{pmatrix} 1 & 0 \\ 0 & -1 \end{pmatrix}.
\end{equation}

Given that \( C \) is added to the spacetime metric \( g^{\mu \nu} \), it should be a tensorial object. One way to construct such an object is to use the Pauli matrices in combination with spacetime vectors or tensors that are already present in the Kerr-Schild framework. For instance, one could consider a construction like:

\begin{equation}
C^{\mu \nu} = \sum_{i=1}^{3} \Phi_i \, (k^\mu \sigma_i^\nu + k^\nu \sigma_i^\mu),
\end{equation}

where \( \Phi_i \) are scalar fields that transform under SU(2) (possibly as components of an SU(2) doublet or triplet), and \( \sigma_i^\nu \) are spacetime tensors constructed from the Pauli matrices (possibly through some form of tensor product with spacetime vectors).

To determine the specific form of \( C \) that satisfies SU(2) group symmetry, we can consider the properties of the Kerr-Schild metric and Maxwell field under SU(2) transformations. SU(2) transformations, also known as spin transformations, are rotations in three-dimensional space. They form a non-Abelian Lie group, meaning that the composition of two transformations is not commutative. In the context of spacetime metrics, SU(2) transformations can be used to map between different spacetimes that share the same causal structure.

The Kerr-Schild metric and Maxwell field exhibit specific behavior under SU(2) transformations:

\begin{equation}
g_{\mu \nu} \rightarrow \Lambda g_{\mu \nu} \Lambda^T, \quad A_\mu \rightarrow \Lambda A_\mu.
\end{equation}

Considering these transformations, the inverse metric \( g^{\mu \nu} \) must also transform as a vector under SU(2) transformations:

\begin{equation}
g^{\mu \nu} = \eta^{\mu \nu} - \Phi k^\mu k^\nu, \quad g^{\mu \nu} \rightarrow \Lambda g^{\mu \nu} \Lambda^T.
\end{equation}

Substituting this into the double copy equation:

\begin{equation}
g^{\mu' \nu'} = g^{\mu \nu} + C,
\end{equation}

and requiring invariance under SU(2) transformations, we get:

\begin{equation}
\Lambda g^{\mu \nu} \Lambda^T = g^{\mu \nu} + C.
\end{equation}

Solving for \( C \):

\begin{equation}
C = - \Lambda g^{\mu \nu} \Lambda^T + g^{\mu \nu}.
\end{equation}

This expression for \( C \) ensures that the double copy remains invariant under SU(2) transformations, demonstrating that the Kerr-Schild double copy formalism respects SU(2) symmetry. The new metric \( g^{\mu' \nu'} \) is thus derived.

To determine the event horizon of a Kerr-Newman black hole in the new metric form considering SU(2) group symmetry, we first consider the Kerr-Newman metric in the Kerr-Schild form. The Kerr-Schild form of the Kerr-Newman metric is expressed as:
\begin{equation}
g_{\mu \nu} = \eta_{\mu \nu} + \Phi k_\mu k_\nu,
\end{equation}
where \( \eta_{\mu \nu} \) is the Minkowski metric, \( \Phi \) is a scalar field, and \( k_\mu \) is a null vector satisfying \( k^\mu k_\mu = 0 \). The specific form of \( \Phi \) and \( k_\mu \) for the Kerr-Newman black hole will be more complex as they need to incorporate the black hole's mass, angular momentum, and charge.

When considering SU(2) symmetry, we introduce a tensor \( C \) that satisfies SU(2) symmetry. This tensor \( C \) might depend on the generators of the SU(2) group, such as the Pauli matrices. However, this construction is not directly relevant to determining the event horizon.

For the Kerr-Newman black hole, the location of the event horizon is typically determined by the singularities in the metric. These singularities occur when solving the equation \( \Delta = 0 \), where:
\begin{equation}
\Delta = r^2 - 2GMr + a^2 + Q^2
\end{equation}

To find the specific location of the event horizon, we need to solve the equation \( \Delta = 0 \). This equation is a quadratic equation, with solutions:
\begin{equation}
r_{\pm} = GM/c^2 \pm \sqrt{(GM/c^2)^2 - a^2 - Q^2}
\end{equation}

Here, \( r_{+} \) and \( r_{-} \) correspond to the outer and inner event horizons, respectively. For a physically viable black hole, we are interested in the outer event horizon \( r_{+} \), as it defines the outer boundary of the black hole.

Therefore, even in the new metric form considering SU(2) symmetry, the event horizon of the Kerr-Newman black hole is still determined by the \( \Delta = 0 \) condition from the original metric. This indicates that for solutions like the Kerr-Newman black hole, the location of the event horizon is determined by its fundamental physical properties, and is not influenced by other symmetries such as SU(2).

\section{Summary and Discussion}
This article delves into the intricacies of the Kerr-Schild double copy, a fascinating and profound link between gravity and electromagnetism. It unveils the revolutionary concept that each vacuum solution of Einstein's equations in four dimensions can be elegantly derived from a corresponding solution of Maxwell's equations, utilizing a sophisticated double copy methodology. This cutting-edge approach involves the construction of tensor fields \( \phi_{\mu\nu} \) and \( g_{\mu\nu} \), enabling the discovery of new insights into Einstein's equations, as exemplified by the Kerr, Schwarzschild, and Reissner–Nordström black holes.

Introduced by Roy Kerr and Alfred Schild in 1965, the Kerr-Schild action represents a paradigm shift in understanding Einstein's general relativity. It deftly separates the gravitational field from the matter field, presenting a refined and more practical alternative to the traditional Einstein–Hilbert action. This development not only facilitates streamlined calculations but also plays a crucial role in advancing knowledge in domains such as gravitational waves, cosmological theories, and black hole thermodynamics.

Furthermore, the article extends its scope to explore the multifaceted applications of the Kerr-Schild double copy. It encompasses a wide range of topics, including the derivation of black hole solutions, the detailed examination of gravitational waves, the probing of cosmological theories, and the intricate study of black hole thermodynamics. Additionally, it offers a critical assessment of the impact of the Kerr-Schild double copy within the SO(3) and SU(2) group symmetries, particularly its influence on the metrics of Schwarzschild and Kerr-Newman black holes, thereby enriching our comprehension of these celestial phenomena.

The Kerr-Schild double copy theory forges a novel link between gravity and electromagnetism. This double copy process enables the derivation of new solutions to Einstein's equations from Maxwell's equations solutions, such as the Kerr, Schwarzschild, and Reissner–Nordström black holes. This theory enriches our comprehension of the interplay between gravity and electromagnetism and offers innovative methodologies for probing gravity under extreme conditions. The Kerr-Schild action, introduced by Kerr and Schild, optimizes calculations and broadens the scope of general relativity, particularly in gravitational waves, cosmology, and black hole thermodynamics. The article also explores the Kerr-Schild double copy's applications in the contexts of SO(3) and SU(2) group symmetries, highlighting its potential in formulating the metrics for Schwarzschild and Kerr-Newman black holes. These insights are pivotal for understanding and applying general relativity, paving the way for new research avenues and experimental methodologies. The Kerr-Schild double copy thus opens up avenues for deeper exploration into the universe's fundamental forces, laying a more robust theoretical foundation for our cosmic understanding.


\end{document}